# Toward Large-Scale Agent Guidance in an Urban Taxi Service

**Lucas AGUSSURJA**
School of Information Systems
Singapore Management University
80 Stamford Road, S178902

**Hoong Chuin LAU**
School of Information Systems
Singapore Management University
80 Stamford Road, S178902

## Abstract

Empty taxi cruising represents a wastage of resources in the context of urban taxi services. In this work, we seek to minimize such wastage. An analysis of a large trace of taxi operations reveals that the services' inefficiency is caused by drivers' greedy cruising behavior. We model the existing system as a continuous time Markov chain. To address the problem, we propose that each taxi be equipped with an intelligent agent that will guide the driver when cruising for passengers. Then, drawing from AI literature on multi-agent planning, we explore two possible ways to compute such guidance. The first formulation assumes fully cooperative drivers. This allows us, in principle, to compute system-wide optimal cruising policy. This is modeled as a Markov decision process. The second formulation assumes rational drivers, seeking to maximize their own profit. This is modeled as a stochastic congestion game, a specialization of stochastic games. Nash equilibrium policy is proposed as the solution to the game, where no driver has the incentive to singly deviate from it. Empirical result shows that both formulations improve the efficiency of the service significantly.

## 1 INTRODUCTION

Taxis are a major mode of transport in every urban city in the world. In Singapore, as of April 2009, there were about 24,000 taxis and 87,000 licensed drivers, providing around 850,000 trips daily. These are operated by a small number of companies. The largest company, ComfortDelgro, operates over 15,000 taxis, and captures the majority of the market share in terms of ridership. Like many congested cities in the world, commuters in Singapore often view taxis as a more efficient mode of transport compared to private cars. Data from the Singapore Land Transport Authority (the regulatory agency for land transportation) show that taxis on average chalk up higher mileage than private cars, with single-shift taxis traveling some 120,000 km a year. More than a third of this travel is empty cruising, as shown in the next section, see (Tan et al., 2009) as well, which represents a significant wastage of resources.

Taxi services have been studied extensively in the literature. Research has been conducted to investigate the services' demand-supply interaction and the resulting market equilibrium (Cairns and Liston-Heyes, 1996; Yang et al., 2002). These studies aim to predict the economic consequences of regulatory policies, such as entry restriction and fare control. At the operational level, quantitative models have been built to capture, for example, passengers' and drivers' bilateral searching behavior (Wong et al., 2005). While these models tend to be descriptive, our work is operative by nature. We seek to provide solutions to a specific problem, namely, we improve the inefficiency of existing cruising policy by providing alternative policies derived from multi-agent planning and decision making models. We believe the public transport arena offers a rich domain for application of multi-agent concepts and methodologies.

We start by analyzing a dataset obtained from a major taxi operator, which traces the movement of a large number of taxis in Singapore. Using the data for the month of July 2009, our analysis shows that the current system's inefficiency, measured in terms of cruising hours, is due to the inherently greedy behavior of drivers. This, in turn, is caused by the lack of visibility in regards to the distribution of cruising taxis on the network at different time periods. We then model the existing taxi service by a continuous-time Markov chain, whose parameters are derived from the dataset. The result is used as the baseline for empirical com-parison with our approach.

We proceed to propose two models: a cooperative and a non-cooperative model. In our formulations, each driver is endowed with a guidance agent that provides suggestions on how to cruise in search of passengers. In the first formulation, we assume that each driver is fully cooperative, and therefore willing to follow a centralized policy (that could be suboptimal for him/her individually). This allows us, in principle, to compute the system-wide optimal policy, where the objective is to maximize overall occupied time. For this purpose, we model the problem as a Markov decision process. In the second formulation, we assume that drivers are rational who seek to maximize their respective occupied time. Here, there is an implicit competition among drivers, simply because an increase in the number of cruising taxis in a zone decreases each driver's chance of finding passengers within a given time. This leads directly to a game theoretic formulation. We model the problem as a stochastic congestion game (a specialization of stochastic games) and seek to find a Nash equilibrium policy. Given an equilibrium policy, no driver has the incentive to singly deviate from it. Both problems are solved for finite horizon using value iteration coupled with a sampling technique.

There has been a surge of interest, in recent years, among AI and complexity theory communities in computational problems related to game theory. The central problem in the area is the computation of Nash equilibrium in different game settings. Classical algorithms to solve the standard simultaneous games have relied on homotopy methods, which solve fixpoint (one instance of which is the Nash equilibrium) problems. The most well-known of such methods is the Lemke-Howson algorithm. See (Herings and Peeters, 2009) for a recent survey, and (Goldberg et al., 2011) for a discussion on the complexity of such methods. Non-homotopy attempts in the literature include enumeration of strategy supports (Porter et al., 2004) and mixed-integer programming (Sandholm et al., 2005).

In the setting of dynamic games, specifically in stochastic games (Shapley, 1953), the problem of computing Nash equilibrium has been cast in the context of multiagent reinforcement learning, first introduced by (Littman, 1994), who shows the convergence of value iteration in 2-player zero-sum stochastic games. (Hu and Wellman, 2003) attempts to generalize the result to $n$-player general-sum games by extending Q-learning algorithm. Their algorithm converges to Nash equilibrium in a restrictive case. Recent attempt by (Kearns et al., 2000) proposes an algorithm that converges to an equilibrium-like joint policy but not Nash equilibrium. Finding an algorithm that converges to Nash equilibrium in $n$-player general-sum stochastic games remains an important open problem.

Despite the tremendous interest, many interesting real-world problems (such as the one presented here) are so large that even the best algorithms have no hope of computing an equilibrium directly. Furthermore, it has been shown that the problem is likely to be intractable even for the case of two-player simultaneous games (Chen and Deng, 2006), it being complete for the complexity class PPAD. The standard approach to overcome this problem is to construct a smaller game that is similar to the original game, and solve the smaller game. Then, the solution is mapped to a strategy profile in the original game (Ganzfried and Sandholm, 2011).

In this work, we tackle a moderate-size real-world problem, by solving for finite horizon approximate equilibrium. Our aim is to extend this solution approach to cover the full scale problem in the future. Our experimental results show that the optimal policy (derived from the cooperative model) manages to reduce the cruising time of the existing policy by approximately 30%. This result may be viewed as what can be achieved theoretically. In real-life implementation, however, the improvement will be closer to that of equilibrium policy, which is approximately 20%.

## 2 ANALYSIS OF CURRENT SYSTEM

Figure 1 and 2 summarize the taxi operation on weekdays (Mon-Thu) for the month of July 2009. In Figure 1a, we see the comparison of average daily time spent between delivering passengers, cruising, and picking up passengers (for booked trips) for different periods of the day. The average number of operating taxis is also shown in the figure. To book a taxi, passengers may send their request to the central operator and specify their current location. The central operator will then broadcast the request to cruising taxis around the specified location. Drivers who receive the request may bid for the job by specifying the time required to reach the pickup point. The job is given to the driver with the shortest pickup time. The proportion of street-hail vs. booked trips, for different periods of the day, is shown in Figure 1b. From Figure 1a, we observe that taxi cruising time is almost constant throughout the day, accounting for roughly half of the total operating hours. On the other hand, pickup time accounts for only a small percentage of operating hours. For this reason, in this work, we are focusing on reducing the cruising time for street-hail jobs.

Figure 2 describes the cruising behavior of drivers. In

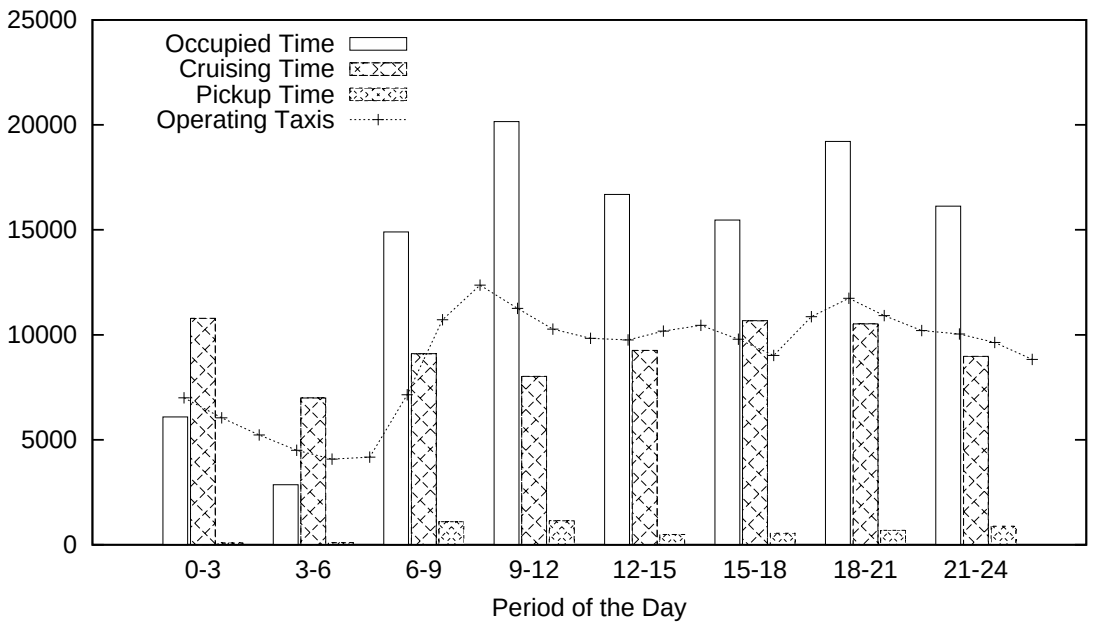

(a) The histogram shows the average total time spent daily, in hours, on: delivering passengers, cruising, and reaching pickup points (for booked trips) respectively, in different periods of the day. The curve shows the number of operating taxis.

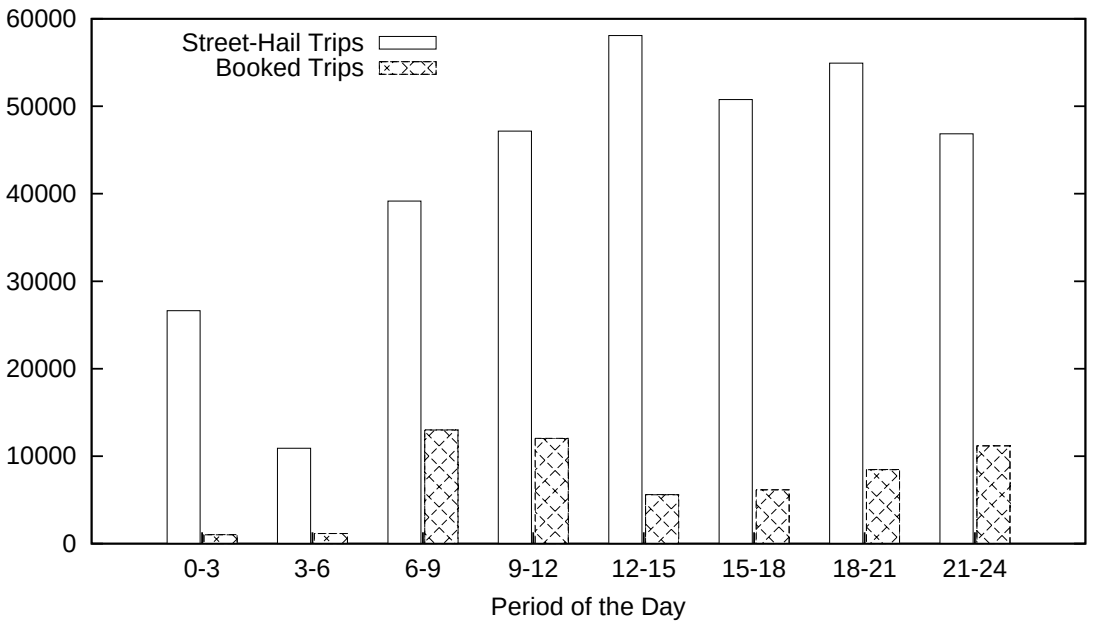

(b) The histogram shows the average daily number of street-hail and booked trips respectively in different periods.

Figure 1: Summary of weekday data for July 2009

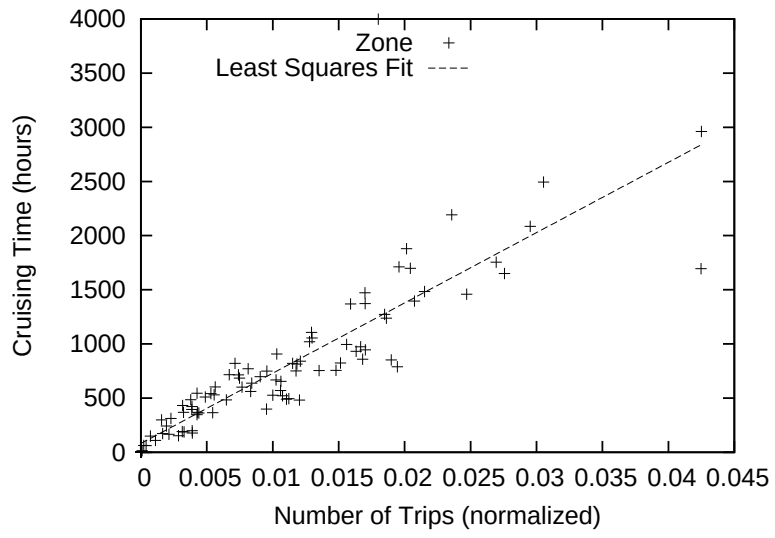

(a) The correlation between trip frequency and cruising time in a zone

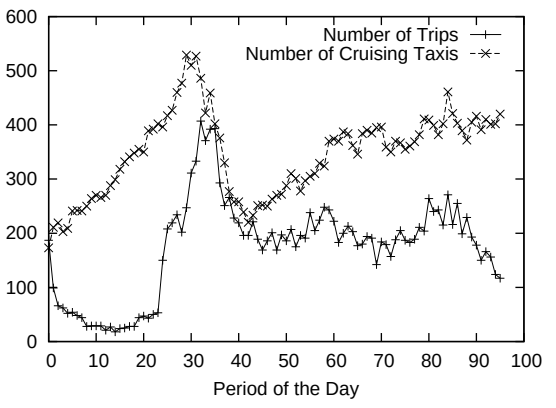

(b) Number of trips vs. cruising taxis in a high trip-frequency zone with peaks

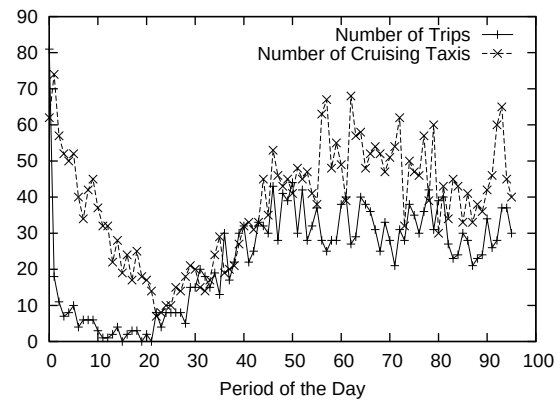

(c) Number of trips vs. cruising taxis in a low trip-frequency zone

Figure 2: Inefficiency of existing cruising policy

Figure 2a, we see that cruising time is high for zones with a high number of available trips. This indicates that drivers are spending time cruising in high trip-frequency zones. The same can be observed from Figure 2b and 2c. Figure 2b shows that taxis are entering high trip-frequency zones in anticipation for the surge of passengers. The number of cruising taxis will drop with the surge, but starts to build up again for the next surge, and the cycle continues. On the other hand, in low trip-frequency zones, except for early morning hours, the number of cruising taxis stays proportional to trip-frequency. We can conclude that drivers are employing a greedy cruising policy, spending a large amount of time cruising in high trip-frequency zone. This is one of the causes for the system's inefficiency, which is verified in the empirical study, when this policy is substituted for better cruising policies.

## 2.1 A MODEL FOR THE EXISTING SYSTEM

In this section, we model the aggregate behavior of a taxi service as a continuous time Markov chain. In our model, the taxi service operates on a road network which can be divided into logical cruising zones. We denote the network of zones by a directed graph $G = (N, E)$. At any point of time, a taxi is in one of the following states, which corresponds to the states in the Markov chain $\mathcal{S} = \{\mathcal{O}_{kl}, \mathcal{C}_k, \mathcal{W}_k | k, l \in N\}$: (1) the taxi is occupied and is delivering its passenger from zone $k$ to $l$, denoted by the state $\mathcal{O}_{kl}$, or (2) the taxi is empty and cruising in a zone $k$, denoted by $\mathcal{C}_k$, or (3) the taxi is in zone $k$ but not in operation, denoted by $\mathcal{W}_k$. Here, we assume that drivers have uniform cruising behavior that are independent of each other. As an example, Figure 3 shows a subset of Singapore's road network and the corresponding Markov chain, modeling the taxi service operating on the network.

In a continuous time Markov chain, the time spent in a state before moving to another state is a continuous random variable that is exponentially distributed. The rates of transition between the states constitute the generator matrix (or $Q$-matrix) of the Markov chain. In our model, the generator matrix consists of the following four components. The first component, $\{\lambda_{kl} | (k, l) \in E\}$, describes a driver's cruising behavior. Since a driver does not have visibility in regards to the state and location of other drivers, we can assume that they are approximately independent of each other. The second component, $\{\pi_{kl} | k, l \in N\}$, describes the likelihood of finding passengers in a zone, where $\pi_{kl}$ is the rate of finding passengers in zone $k$ with a destination point in zone $l$. Assuming independence for $\pi_{kl}$, the total rate of finding passengers in zone $k$ is given by $\sum_l \pi_{kl}$ (a combination of Poisson processes). The third component, $\{\rho_{kl} | k, l \in N\}$, describes the time needed to deliver passengers to their destination point. The nondeterminism of these variables is due to the variability of pickup and drop-off points within zones and the congestion on the road network. The forth component, $\{\sigma_k, \sigma'_k | k \in N\}$, describes the likelihood of a driver taking a break within a zone ($\sigma_k$) and the length of the break ($\sigma'_k$). Most taxis are operated by two or more drivers taking turns continuously. This accounts for the nondeterminism of the forth component.

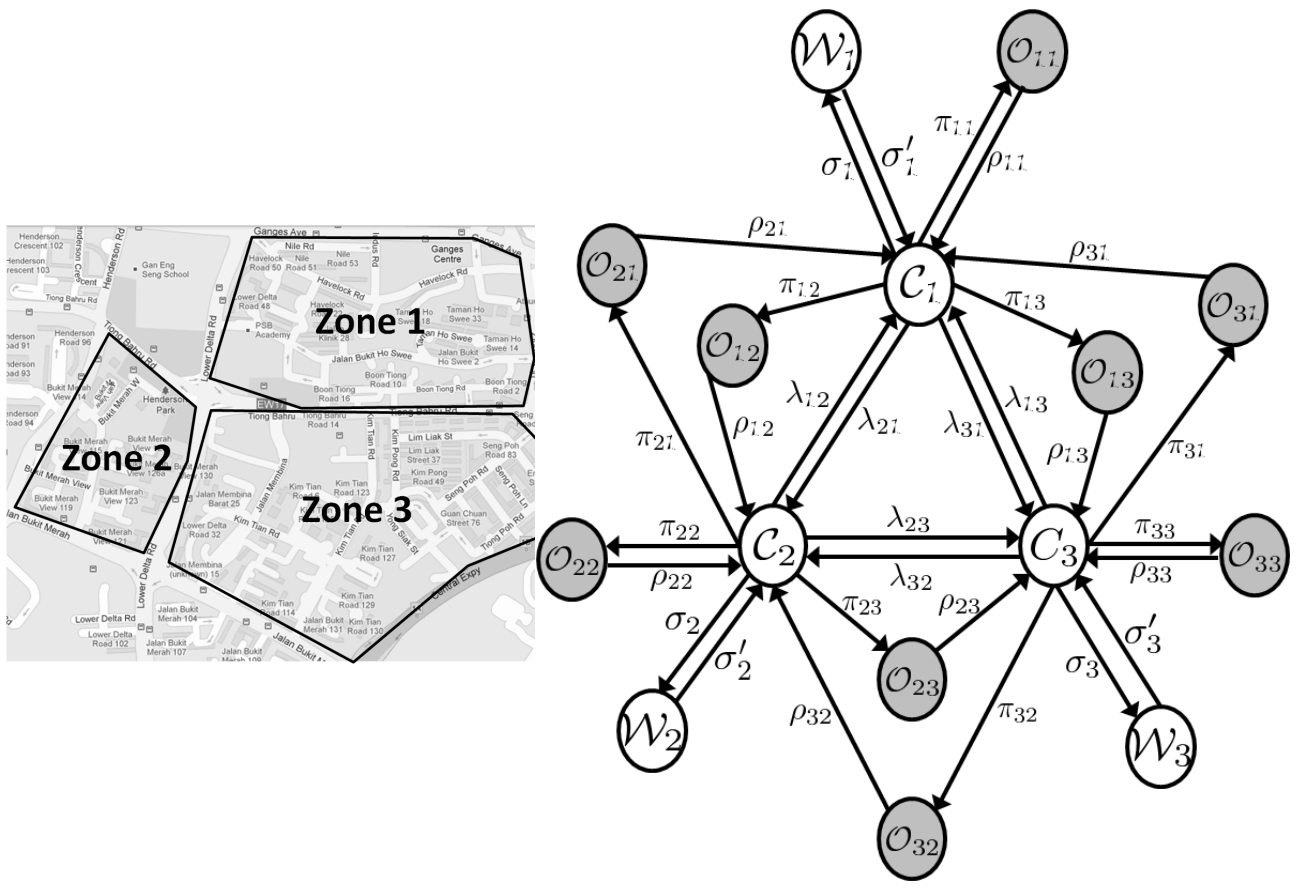

Figure 3: An example of a road network, divided into its cruising zones, and the corresponding Markov chain, modeling a taxi operating on the network. The shaded nodes indicate the states where the taxi is occupied. Here, we assume that taxis are uniform and independent of each other.

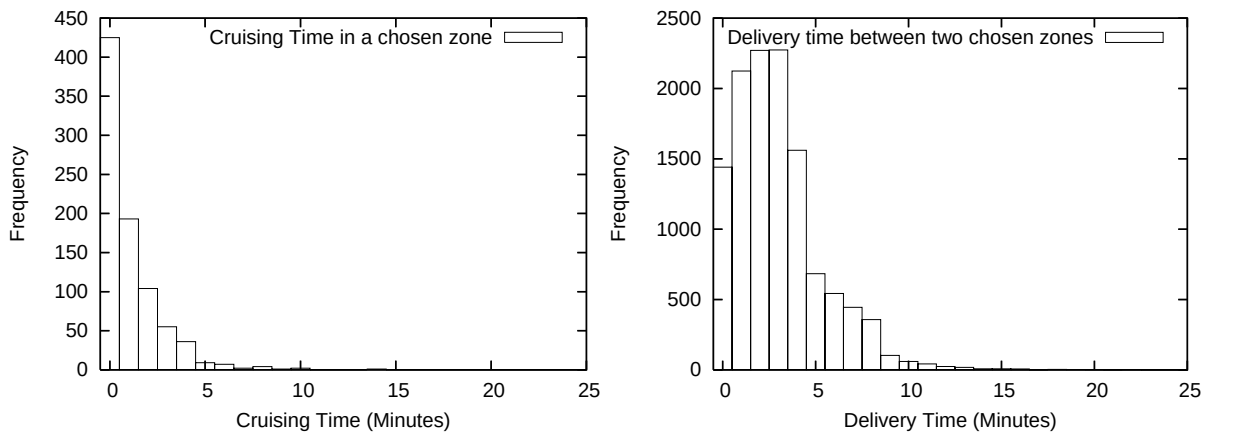

Figure 4: The frequency histograms (for one day) for cruising time in a randomly chosen zone, and delivery time between two randomly chosen zones, respectively.

We define the system efficiency as the steady state (stationary) probability of a taxi being in the occupied state. Let $\theta(s)$, for $s \in \mathcal{S}$, denote the steady state probability of being in the state $s$. This probability distribution can be obtained by solving the following system of equations:

$$\begin{aligned}
\forall k \in N, \quad & \sum_{l:(l,k)\in E} \lambda_{lk}\theta(\mathcal{C}_l) + \sum_{l:l\in N} \rho_{lk}\theta(\mathcal{O}_{lk}) + \sigma'_k\theta(\mathcal{W}_k) \\
& = \left( \sum_{l:(k,l)\in E} \lambda_{kl} + \sum_{l:l\in N} \pi_{kl} + \sigma_k \right) \theta(\mathcal{C}_k), \\
\forall k, l \in N, \quad & \pi_{kl}\theta(\mathcal{C}_k) = \rho_{kl}\theta(\mathcal{O}_{kl}), \\
\forall k \in N, \quad & \sigma_k\theta(\mathcal{C}_k) = \sigma'_k\theta(\mathcal{W}_k),
\end{aligned}$$

and normalizing the solution such that $\sum_{s\in\mathcal{S}} \theta(s) = 1$. The steady state probability of a taxi being in the occupied state is thus given by $\sum_{k,l\in N} \theta(\mathcal{O}_{kl})$.

We estimate the transition rates of the Markov chain using the dataset. The random variables cruising time (with rate $\lambda_{kl}$) and passenger find time (with rate $\pi_k$) can be estimated accurately with exponential distributions. The other variables, however, are closer to Erlang distributions than exponential distributions. Figure 4 shows, for example, two frequency histograms, one for each of these cases. In this work, we approximate both cases by exponential distributions using maximum likelihood estimate. These approximations are better when the zones are adequately close to each other, which is the case for Singapore.

One can derive the system efficiency once the parameters of the Markov chain are obtained. Since we also like to compute efficiency under different customer arrival rates and different numbers of operating taxis (for empirical study), we model the passenger-driver dynamics in a zone as an M/M/1 queue. Consider the subset of the Markov chain consisting only the states $\{\mathcal{C}_k | k \in N\}$ and the transitions rate between them $\{\lambda_{kl} | (k,l) \in E\}$. Each zone is modeled as a FIFO queue, where its arrival rate is the arrival rate of cruising taxis in the zone, and its service rate is the passengers arrival rate in the zone. Let $\mu_k$ denote the passengers arrival rate in zone $k$, then the waiting time in the queue (queueing time + service time) is exponentially distributed with the following rate, which we associate with $\pi_k$:

$$\pi_k = \mu_k - n \left( \sum_{l:(l,k)\in E} \phi(\mathcal{C}_l)\lambda_{lk} \right) \quad ,$$

where $n$ is the number of taxis in operation and $\phi(\mathcal{C}_l)$ the steady state probability of being in the state $\mathcal{C}_l$. The steady state probabilities $\phi$ can be obtained, similar to $\theta$, by solving the following system of equations:

$$\forall k \in N, \quad \sum_{l:(l,k)\in E} \lambda_{l,k}\phi(\mathcal{C}_l) = \phi(\mathcal{C}_k) \sum_{l:(k,l)\in E} \lambda_{kl} \quad ,$$

subject to $\sum_{k\in N} \phi(\mathcal{C}_k) = 1$. Now, given the drivers' cruising behaviors (in the form of smaller Markov chain parameterized by $\{\lambda_{kl} | (k,l) \in E\}$), travel time information $\{\rho_{kl} | k, l \in N\}$, and nonoperating profile $\{\sigma_k, \sigma'_k | k \in N\}$, we can derive the system efficiency as the function of passenger arrival rates $\{\mu_k | k \in N\}$ and the total number of taxis $n$, by first constructing the full Markov chain, and then deriving its steady state probabilities.

# 3 A COOPERATIVE MODEL

The cooperative model is based on Markov decision process, which is a special case of stochastic games, where there is only one player. The player, in this case, is a central operator whose actions are all possible joint actions of the drivers, and whose objective is the overall occupied time of the service. The model is similar to the noncooperative case. The difference lies in the formulation of the utility function. In the noncooperative case, we have a set of utility functions to be optimized simultaneously, while in the cooperative case, we have a single aggregated utility function. We choose to present our models under the more general setting of stochastic games (next section). We will highlight the differences for the cooperative case.

# 4 A NONCOOPERATIVE MODEL

Stochastic games (Shapley, 1953; Littman, 1994) are a generalization of Markov decision processes to a multi-agent setting by allowing the state transitions to be influenced by their joint action. They are also a generalization of sequential games with perfect information, by introducing different states. The state (and thus the payoff matrix) changes as a result of both nature and the joint action of the agents. In congestion games, agents share a set of facilities (facilities can be viewed as resources, such as machines), and the utility an agent derives from using a facility reduces as the number of agents using the same facility increases. In the taxi context, the facilities correspond to the zones where taxi agents can cruise. Each zone has a fixed arrival rate of passengers, and therefore, an increase in the number of cruising taxis in the zone, decreases the likelihood that each would find passengers within a certain time period. Stochastic congestion games therefore are a generalization of congestion games, allowing the games to be played indefinitely. They are also a specialization of stochastic games, to the case where the underlying games are congestion games. A state in a stochastic congestion game defines an assignment of agents to facilities, and transitions to a new state are dependent on the old state and the agents' joint action. In this work, we consider only games with finite horizon.

## 4.1 STOCHASTIC CONGESTION GAMES

Formally, a finite stochastic congestion game (SCG) $\Gamma$ is a tuple $(I, J, P, \mathcal{R})$, where

- $I = \{1, \ldots, n\}$ is the set of agents.
- $J = \{1, \ldots, m\}$ is the set of facilities. An action of an agent $i$, denoted by $a^i$ where $a^i \in J$, is to choose a facility to which it would like to be assigned. We denote an agents' joint action by $a = (a^1, \ldots, a^n)$ and the set of possible joint actions by $A$, that is, $A = J^n$. Next, we denote by $S$, the set of possible states, where a state $s \in S$ is an assignment of agents to facilities, and we define $s_j$ as the number of agents assigned to facility $j$ in the state $s$.
- $P : S \times A \to \triangle(S)$ is the state transition function. For convenience, we will also write $P(s, a, s')$ for the probability that the next state is $s'$ given that the current state is $s$ and the players' joint action is $a = (a^1, \ldots, a^n) \in A$,
- $\mathcal{R} = \{r_j\}_{j \in J}$ is the reward function, where each $r_j : \{0, \ldots, n\} \to \mathbb{R}$ is a nonincreasing function that maps the number of agents assigned to facility $j$ to the reward that each of the agents in $j$ receives. We define $s(i, j)$ such that $s(i, j) = 1$ if agent $i$ is assigned to facility $j$ in the state $s$, and $s(i, j) = 0$ otherwise. The reward received by agent $i$ in a state $s$, denoted by $R^i(s)$, is thus given by $\sum_{j \in J} s(i, j) r_j(s_j)$.

The game proceeds in steps, starting from some initial state $s_T$ (the subscript indicates the number of remaining steps). In each step, the agents first observe the current state $s_t$, the number of remaining steps $t$, and simultaneously choose actions according to their respective policy. An agent $i$'s policy is the function $\pi^i : S \times \{1, \ldots, T\} \to \triangle(J)$, where $\pi^i(s_t, t)$ computes agent $i$'s mixed strategy, i.e., a probability distribution over the set of facilities, given the current state $s_t$ and the number of remaining steps $t$. Each agent $i$'s action in this step, denoted by $a_t^i$, is then drawn from the distribution given by $\pi^i(s_t, t)$, forming the joint action $a_t = (a_t^i, \ldots, a_t^n)$. Nature then selects the next state $s_{t-1}$ according to the probabilities given by $P(s_t, a_t)$. In the new state $s_{t-1}$, each agent $i$ receives $R^i(s_{t-1})$ as its reward, and the game proceeds to the next step.

Given a possible outcome of the game, agent $i$'s utility, denoted by $U^i$, is defined such that

$$U_T^i(s_T, a_T, s_{T-1}, \ldots, s_1, a_1, s_0) = \sum_{t=0}^{T-1} R^i(s_t).$$

A Nash equilibrium of the game is a joint policy $\pi$, such that, for each agent $i$, $\pi^i$ maximizes the expected utility of agent $i$ given that the other players follow their respective policy specified by $\pi$. The expectation is taken over all possible outcomes of the game. For the cooperative case, there is only one utility formed by summing up individual agent's utility. Here, we seek to find a joint policy that maximizes the expected value of this utility.

### 4.2 MODELING TAXI SYSTEM AS AN SCG

The drivers correspond to the set of agents $I$. The set of facilities $J$ corresponds to the states of the Markov chain, $\mathcal{S} = \{\mathcal{C}_k, \mathcal{O}_{kl}, \mathcal{W}_k | k, l \in N\}$, defined previously. We will refer to them as facilities here. A state of the game is an assignment of agents to facilities. When an agent is assigned to facility $\mathcal{C}_k$, it means that the corresponding taxi is cruising in zone $k$. Similarly, when it is assigned to facility $\mathcal{O}_{kl}$ and $\mathcal{W}_k$, it means that the corresponding taxi is occupied and not in operation, respectively. One step in the SCG represents the period of one minute.

Next, we define available actions and the state transition function. The set of available actions for an agent depends on the state of the agent. When in facility $\mathcal{C}_k$, the agent may take one of the following actions: $(a1)$ continue cruising in the current zone, or $(a2)$ make an attempt to move to an adjacent zone. In this state, the agent has the chance of getting a passenger and be moved to one of the facilities $\mathcal{O}_{kl}$ in the next step. Given that an agent is in facility $\mathcal{C}_k$ and takes the action $a1$, the following may happen in a step: (1) The agent manages to find a passenger with $l$ as the destination zone. In the next step, the agent will move to facility $\mathcal{O}_{kl}$. The probability of this event happening is given by:

$$\frac{\mu_k}{n(\mathcal{C}_k)}\gamma_{kl},$$

where $\gamma_{kl}$ is the probability that a passenger's destination zone is $l$ given that its starting zone is $k$, and $n(\mathcal{C}_k)$ is the number of agents in facility $\mathcal{C}_k$ in the current state. $\gamma_{kl}$ can be estimated by the following:

$$\gamma_{kl} \approx \frac{\pi_{kl}}{\sum_{l \in N} \pi_{kl}}.$$

The expression $\mu_k/n(\mathcal{C}_k)$ also defines the reward of the agent in this state. (2) The agent doesn't find any passenger and stays in the same facility. The probability of this event is given by: $(1 - \mu_k/n(\mathcal{C}_k))e^{-\sigma_k}$, or (3) the agent doesn't find any passenger and moves to facility $\mathcal{W}_k$ with probability $(1 - \mu_k/n(\mathcal{C}_k))(1 - e^{-\sigma_k})$. On the other hand, if the agent chooses $a2$, one of the following may occur: (1) a passenger, with destination $l$, is found before the agent manages to move to a new zone. The agent moves to facility $\mathcal{O}_{kl}$ in the next step. The probability of this happening is given by $\mu_k \gamma_{kl}/n(\mathcal{C}_k)$. (2) Otherwise, no passenger is found, and the agent moves to a new zone.

**Algorithm** FINITENASHPOLICIES$(\Gamma, T)$:
**Initialization phase:**
  For all $s \in S$, joint action $a \in A$:
    $\pi(s, 0)[a] \leftarrow \frac{1}{|A|}$;
    For all $i \in I$:
      $V_{s,0}(a, i) \leftarrow 0$;
**Iteration phase:**
  For $t = 1$ to $T$, all $s \in S$:
    For all pure strategy profile $a \in A$, $i \in I$:
      $V_{s,t}(a, i) \leftarrow \sum_{s'} P(s, a, s') \times$
      $\{R^i(s') + \sum_{a'} \pi(s', t-1)[a'] V_{s',t-1}(a', i)\}$;
    $\pi(s, t) \leftarrow$ FINDNASH$(V_{s,t})$;
  Return $\pi$;

Figure 5: Value Iteration for Finite-horizon SCGs

When in facility $\mathcal{O}_{kl}$, the agent has only one available action. It will try to deliver its passengers to their destination point and move to facility $C_l$. The probability of this happening in the next step is given by $1 - e^{-\rho_{kl}}$, while the probability of staying in the same facility is $e^{-\rho_{kl}}$. The reward for the agent in this state is one. Similarly, when in facility $\mathcal{W}_k$, the agent may move to $\mathcal{C}_k$ with probability $1 - e^{-\sigma'_k}$, or otherwise stays in the same zone. The reward for being in this facility is zero. Note that in our model, the move to facility $\mathcal{W}_k$ is involuntary on the part of the agents. An agent will never choose to stop its operation because of the way the reward function is structured. Rather than being an available action, the move is treated as a requirement instead, for example, when the driver has to change shift or rest.

### 4.3 COMPUTING EQUILIBRIUM POLICY

In this section, we describe an algorithm to computing equilibrium policy in SCGs based on value iteration algorithm, shown in Figure 5. This algorithm takes an SCG $\Gamma$, and a time horizon $T$ as input. It outputs a vector of policies $(\pi^i(s,t))_{i \in I}$, where each $\pi^i$ maps any state $s$, and the number of remaining steps $t \leq T$ to a mixed strategy for player $i$, which is a probability distribution over the set of facilities. The algorithm outputs a vector of policies that is a Nash equilibrium for the $T$-step SCG $\Gamma$ from any start state. This algorithm is a generalization of the classical finite-horizon value iteration for Markov decision processes (Kaelbling et al., 1998). Instead of backup values, the algorithm maintains backup matrices, denoted by $V_{s,t}$. Each $V_{s,t}(a, i)$ is a function that takes a joint action $a$, a player $i$, and returns the expected utility of player $i$ in the $t$-step SCG $\Gamma$, if the game starts at the state $s$ and the first joint action of the agents is $a$. We can view, therefore, each $V_{s,t}$ as defining a game in the usual sense, and associated with each $V_{s,t}$ is a Nash equilibrium $\pi(s,t)$, where $\pi^i(s,t)$ denotes the mixed strategy of player $i$. We denote by $\pi(s,t)[a]$ the probability assigned by $\pi(s,t)$ to the joint strategy $a$. Both $V_{s,t}$ and $\pi(s,t)$ are constructed iteratively. The algorithm uses the function FINDNASH that returns a Nash equilibrium given the game $V_{s,t}$. For the cooperative case, the function FINDNASH is replaced by FINDOPT which computes social optimum policy instead of Nash equilibrium.

Computing Nash equilibrium in stochastic games is a very difficult problem. See, for example, (Bowling, 2000; Hu and Wellman, 2003; Ganzfried and Sandholm, 2009) for recent works in this area. And the value iteration algorithm (for finite horizon stochastic games) is computationally very expensive for large scale problem. In this work, we employ the sparse sampling technique proposed by (Kearns et al., 2000).

# 5 EMPIRICAL RESULT

We run a simulation (on a macro level) to evaluate the performance of our proposed solutions. We choose, as the case study, 15 connected zones that represent the congested central business district of Singapore and its surrounding areas. We consider only passengers with origins and destinations within these zones, and restrict the cruising area of drivers to these zones as well. The number of operating taxis is 500. From the dataset we estimate the the passenger arrival rates $\{\mu_k | k \in N\}$, delivery time rate $\{\rho_{kl} | k, l \in N\}$, and taxis nonoperating profile $\{\sigma_k, \sigma'_k | k \in N\}$ for these zones in different periods. Using this setting we compare existing cruising policy (also derived from the dataset), the optimal policy (computed from the cooperative model), and the equilibrium policy (computed from the noncooperative model). The measure of comparison is cruising time. For both cooperative and noncooperative model, joint actions are computed every minute with 2-step look ahead ($T = 2$).

We simulate one day of operation on a weekday. The driver-passenger dynamics in each zone is implemented as a FIFO queue. As a passenger appears in a zone, it joins the queue associated with the zone. If there are some taxis cruising in the zone, the passenger at the head of the queue is removed and assigned to a randomly chosen taxi. The average time spent in the queue models the average passengers waiting time.

The experimental results are shown in Figures 6, 7 and 8. Figure 6 compares the average cruising time (over multiple runs) of the three policies. On average, the optimal policy and the equilibrium policy reduce the cruising time of existing policy by approximately 30% and 20% respectively. Most of the savings are obtained during the peak periods. This matches our intuition, that the new policies are able to better distribute cruising taxis among the zones, while the greedy policy sends too many cruising taxis to high trip-frequency zones. This is confirmed when we look at the number of trips vs. cruising taxis in both high and low trip-frequency zone (see Figure 7 and 8). The new policies are able to maintain balance between fulfilling trips requirement and managing cruising time.

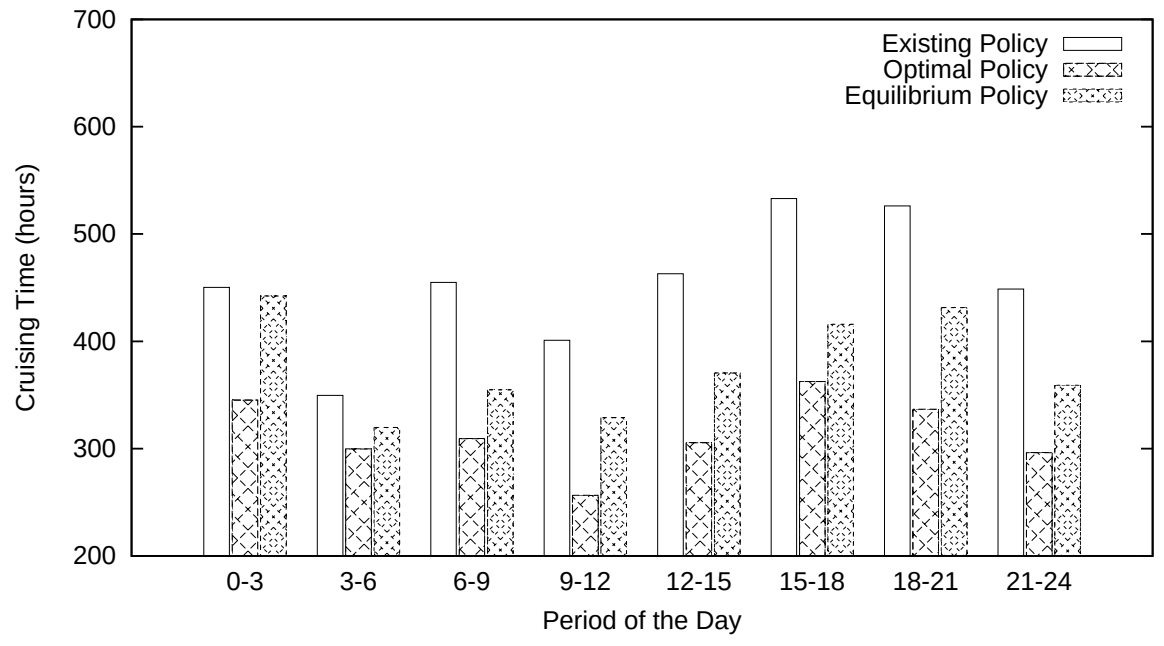

Figure 6: Comparison of cruising time between three policies. The optimal and equilibrium policies reduce the cruising time by approximately 30% and 20% respectively.

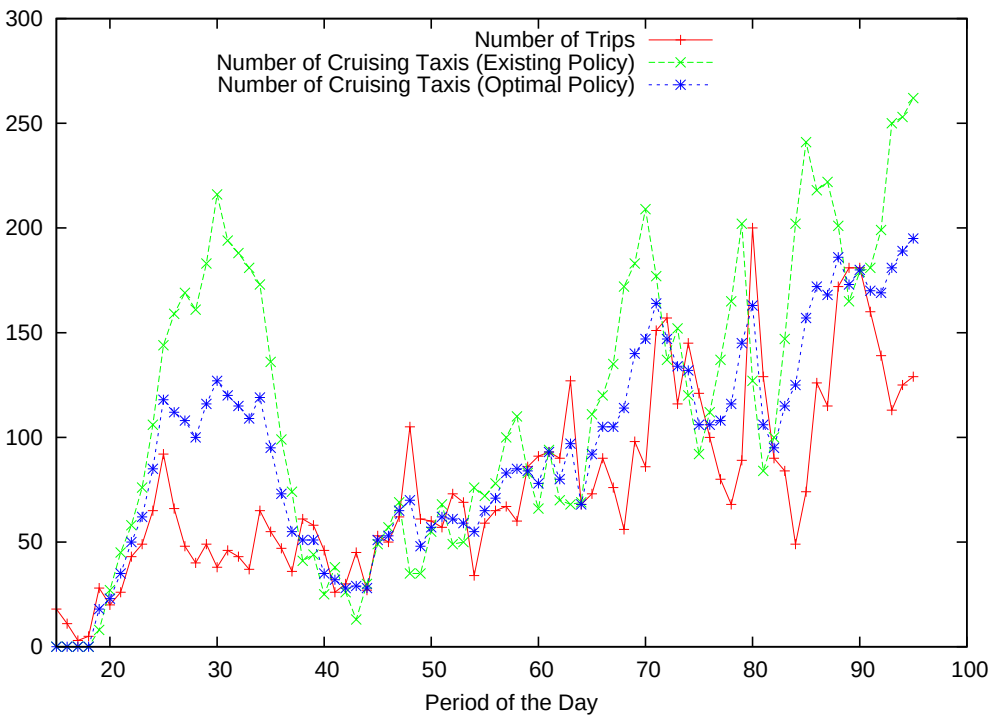

Figure 7: Number of trips vs. number of cruising taxi in a high trip-frequency zone. In general, the new policies send lower number of cruising taxis to high trip-frequency zones compared to existing policy.

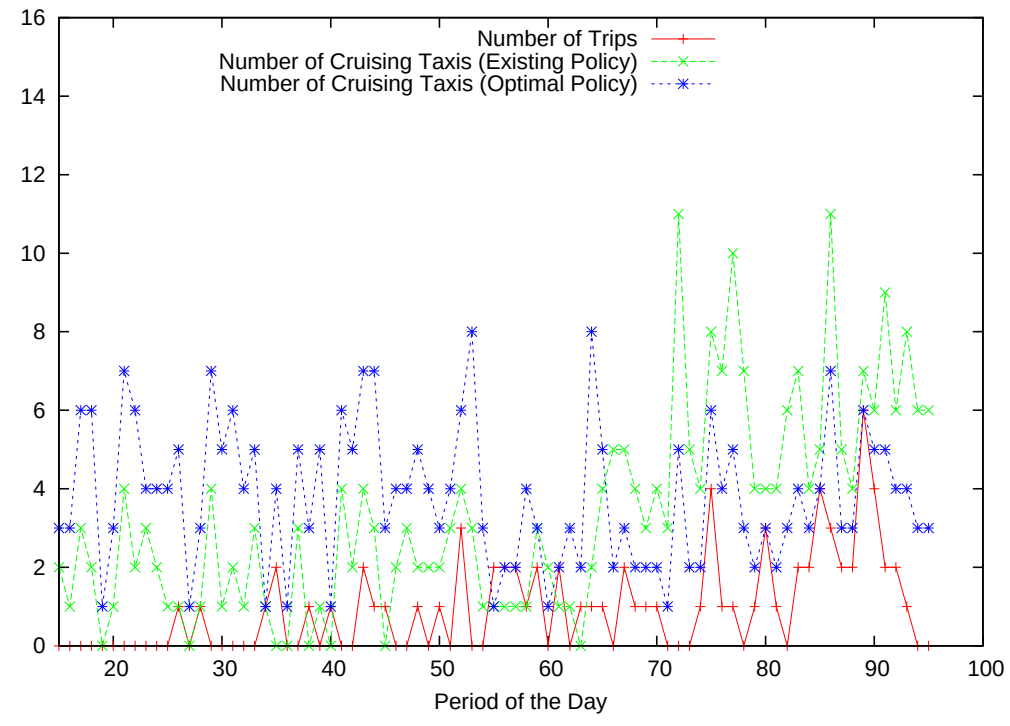

Figure 8: Number of trips vs. number of cruising taxi in a low trip-frequency zone. In general, the new policies send higher number of cruising taxis to low trip-frequency zones compared to existing policy.

## 6 Conclusion

We presented in this paper an interesting and useful application of Markov chains to manage an urban taxi service. The basic premise is the need to minimize cruising time (and therefore maximize utilization) of the taxis. There are several assumptions we made for this to work. We assume that each taxi has a device that guides the taxi driver. Using MDPs the system then generates a policy that optimizes cruising among all taxis. We showed, with the use of real data from a taxi operator in Singapore for the study, a cooperative model where taxis follow instructions and a noncooperative one where drivers compete with one another. Our study also assumes rational drivers that try to maximize the time their cars are occupied. For future works, we would need to incorporate real-world behavior of the passengers as well as relax the independence assumption of passengers and their destinations.